# Endlessly single-mode holey fibers: the influence of core design

**Kunimasa Saitoh, Yukihiro Tsuchida, and Masanori Koshiba**

*Division of Media and Network Technologies, Hokkaido University, Sapporo 060-0814, Japan*
*ksaitoh@ist.hokudai.ac.jp*

**Niels Asger Mortensen**

*MIC - Department of Micro and Nanotechnology, NanoDTU*
*Technical University of Denmark, DK-2800 Kongens Lyngby, Denmark*
*nam@mic.dtu.dk*

**Abstract:** In this paper we evaluate the cut-off properties of holey fibers (HFs) with a triangular lattice of air holes and the core formed by the removal of a single (HF1) or more air holes (HF3 and HF7). With the aid of finite-element simulations we determine the single-mode and multi-mode phases and also find the air hole diameters limiting the endlessly single-mode regime. From calculations of $V$ and $W$ parameters we find that in general HF1 is less susceptible to longitudinal non-uniformities compared to the other designs for equivalent effective areas. As an example we illustrate this general property for the particular case of a macro-bending induced loss.

## References and links

1. J. C. Knight, T. A. Birks, P. S. Russell, and D. M. Atkin, "All-silica single-mode optical fiber with photonic crystal cladding," Opt. Lett. **21**, 1547-1549 (1996).
2. T. A. Birks, J. C. Knight, and P. S. Russell, "Endlessly single-mode photonic crystal fiber," Opt. Lett. **22**, 961-963 (1997).
3. M. D. Nielsen, J. R. Folkenberg, and N. A. Mortensen, "Singlemode photonic crystal fibre with effective area of 600 μm$^2$ and low bending loss," Electron. Lett. **39**, 1802-1803 (2003).
4. M. D. Nielsen, N. A. Mortensen, M. Albertsen, J. R. Folkenberg, A. Bjarklev, and D. Bonacinni, "Predicting macrobending loss for large-mode area photonic crystal fibers," Opt. Express **12**, 1775-1779 (2004), http://www.opticsexpress.org/abstract.cfm?URI=OPEX-12-8-1775.
5. M. D. Nielsen, J. R. Folkenberg, N. A. Mortensen, and A. Bjarklev, "Bandwidth comparison of photonic crystal fibers and conventional single-mode fibers," Opt. Express **12**, 430-435 (2004), http://www.opticsexpress.org/abstract.cfm?URI=OPEX-12-3-430.
6. J. R. Folkenberg, M. D. Nielsen, N. A. Mortensen, C. Jakobsen, and H. R. Simonsen, "Polarization maintaining large mode area photonic crystal fiber," Opt. Express **12**, 956-960 (2004), http://www.opticsexpress.org/abstract.cfm?URI=OPEX-12-5-956.
7. J. R. Folkenberg, M. D. Nielsen, and C. Jakobsen, "Broadband single-polarization photonic crystal fiber," Opt. Lett. **30**, 1446-1448 (2005).
8. N. A. Mortensen, M. D. Nielsen, J. R. Folkenberg, A. Petersson, and H. R. Simonsen, "Improved large-mode-area endlessly single-mode photonic crystal fibers," Opt. Lett. **28**, 393-395 (2003).
9. J. Limpert, A. Liem, M. Reich, T. Schreiber, S. Nolte, H. Zellmer, A. Tünnermann, J. Broeng, A. Petersson, and C. Jakobsen, "Low-nonlinearity single-transverse-mode ytterbium-doped photonic crystal fiber amplifier," Opt. Express **12**, 1313-1319 (2004), http://www.opticsexpress.org/abstract.cfm?URI=OPEX-12-7-1313.
10. N. A. Mortensen, "Effective area of photonic crystal fibers," Opt. Express **10**, 341-348 (2002), http://www.opticsexpress.org/abstract.cfm?URI=OPEX-10-7-341.
11. B. T. Kuhlmey, R. C. Mcphedran, and C. M. de sterke, "Modal cutoff in microstructured optical," Opt. Lett. **27**, 1684-1686 (2002).

12. K. Saitoh and M. Koshiba, "Full-vectorial imaginary-distance beam propagation method based on a finite element scheme: Application to photonic crystal fibers," IEEE J. Quantum Electron. **38**, 927-933 (2002).
13. F. Brechet, J. Marcou, D. Pagnoux, and P. Roy, "Complete analysis of the characteristics of propagation into photonic crystal fibers, by the finite element method," Opt. Fiber Technol. **6**, 181-191 (2000).
14. M. Koshiba, "Full-vector analysis of photonic crystal fibers using the finite element method," IEICE Trans. Electron. **85-C**, 881-888 (2002).
15. G. Renversez, F. Bordas, and B.T. Kuhlmey, "Second mode transition in microstructured optical fibers: determination of the critical geometrical parameter and study of the matrix refractive index and effects of cladding size," Opt. Lett. **30**, 1264-1266 (2005).
16. N. A. Mortensen, J. R. Folkenberg, M. D. Nielsen, and K. P. Hansen, "Modal cutoff and the V parameter in photonic crystal fibers," Opt. Lett. **28**, 1879-1881 (2003).
17. M. D. Nielsen and N. A. Mortensen, "Photonic crystal fiber design based on the V-parameter," Opt. Express **11**, 2762-2768 (2003), http://www.opticsexpress.org/abstract.cfm?URI=OPEX-11-21-2762.
18. M. Koshiba and K. Saitoh, "Applicability of classical optical fiber theories to holey fibers," Opt. Lett. **29**, 1739-1741 (2004).
19. K. Saitoh and M. Koshiba, "Empirical relations for simple design of photonic crystal fibers," Opt. Express **13**, 267-274 (2005), http://www.opticsexpress.org/abstract.cfm?URI=OPEX-13-1-267.
20. M. Koshiba and K. Saitoh, "Simple evaluation of confinement losses in holey fibers," Opt. Commun. **253**, 95-98 (2005).
21. N. A. Mortensen and J. R. Folkenberg, "Low-loss criterion and effective area considerations for photonic crystal fibres," J. Opt. A-Pure Appl. Opt. **5**, 163-167 (2003).
22. Y. Tsuchida, K. Saitoh, and M. Koshiba, "Design and characterization of single-mode holey fibers with low bending losses," Opt. Express **13**, 4770-4779 (2005), http://www.opticsexpress.org/abstract.cfm?URI=OPEX-13-12-4770.
23. J. Riishede, N. A. Mortensen, and J. Lægsgaard, "A 'poor man's approach' to modelling micro-structured optical fibres," J. Opt. A-Pure Appl. Opt. **5**, 534-538 (2003).
24. N. A. Mortensen, "Semianalytical approach to short-wavelength dispersion and modal properties of photonic crystal fibers," Opt. Lett. **30**, 1455-1457 (2005).
25. J. Olszewski, M. Szpulak, and W. Urbańczyk, "Effect of coupling between fundamental and cladding modes on bending losses in photonic crystal fibers," Opt. Express **13**, 6015-6022 (2005), http://www.opticsexpress.org/abstract.cfm?URI=OPEX-13-16-6015.

## 1. Introduction

Large-mode-area (LMA) single-mode fibers are essential to many applications including high power delivery, fiber amplification, and low non-linear data transmission. While standard optical fibers have difficulties in providing both large mode area and single-mode operation, all-silica photonic crystal fibers [1], also referred to as holey fibers (HFs), are highly potential candidates for realizing LMA single-mode fibers. This owes mainly to their endlessly single-mode properties [2] which allows for very broad-band single mode LMA HFs [3]-[5] and even broad-band polarization maintaining and single-polarizing LMA HFs have been achieved with the addition of stress-applying parts [6], [7].

Typically, HFs are formed by a single missing air hole (HF1) in the otherwise periodic triangular lattice (with pitch $\Lambda$) of air holes (with diameter $d$), but HFs with three (HF3) [8] and seven (HF7) [9] missing neighboring air holes have also been reported, see Table 1. In Ref. [8] the HF3 design was suggested to be less susceptible to longitudinal non-uniformities (including macro-bending induced loss) compared to the HF1 for equivalent effective areas. The comparison relied heavily on the ability to predict the respective values of $d^*/\Lambda$ for the endlessly single-mode regime. Unfortunately, super-cell methods, as employed in Ref. [8], have later proved inadequate for such quantitative predictions (for HF3 a value of 0.25 was estimated) as is evident by comparing the phase diagram for HF1 first suggested by Mortensen [10] to the subsequent results obtained by Kuhlmey *et al.* with the aid of a multi-pole method [11]. In this work we employ a finite-element approach which, combined with perfectly-matched layers, also provides quantitative correct results for the cut-off properties.

Our results for HF3 differ significantly from the initial results presented in Ref. [8] and in general we find that the HF1 is indeed superior to the HF3 and HF7 in terms of the susceptibility to longitudinal non-uniformities for equivalent effective areas. We illustrate this general property by means of calculations of the *V* and *W* parameters and we also illustrate it for the particular case of a macro-bending induced loss mechanism.

Table 1. Table of central fiber parameters.

| | HF1 | HF3 | HF7 |
|---|---|---|---|
| | 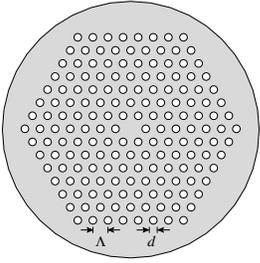 | 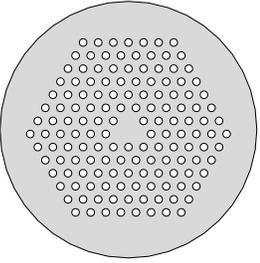 | 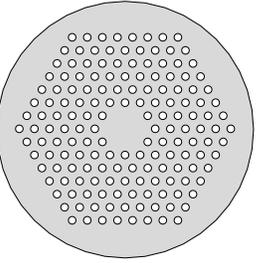 |
| $a_{eff}/\Lambda$ | $1/\sqrt{3}$ | 1 | $\sqrt{2}$ |
| $d^*/\Lambda$ | 0.424 | 0.165 | 0.046 |

## 2. Endlessly single-mode structure and the V parameter

Following the ideas in Ref. [10] we map the cut-off properties in a $\lambda_c/\Lambda$ versus $d/\Lambda$ phase-diagram for the three HFs in Table 1. For the simulations we have employed a finite-element method (FEM) with perfectly-matched layers [12]. In Fig. 1 we summarize the results obtained by the FEM analysis (the open data points) for the three classes of HFs. The cutoff wavelength of the second order mode is defined to be the wavelength at which the effective index of the second order mode becomes equal to the effective cladding index, i.e. the effective index of fundamental space-filling mode, $n_{FSM}$. The accurate value of $n_{FSM}$ can be determined by applying the full-vector FEM to the so-called elementary piece in the cladding region, which acts as a boundary-less propagation medium [13], [14]. A reduced value of $d/\Lambda$ is needed to keep the holey fiber with the inclusion of more than a single solid rod endlessly single mode. For the single rod case the limit is $d^*/\Lambda=0.424$. This value is in excellent agreement with the result of multipole method [15]. For the other holey fiber cores formed by three and seven neighboring rods, we found, respectively, $d^*/\Lambda=0.165$ and $d^*/\Lambda=0.046$ as the upper limits for endlessly single-mode operation.

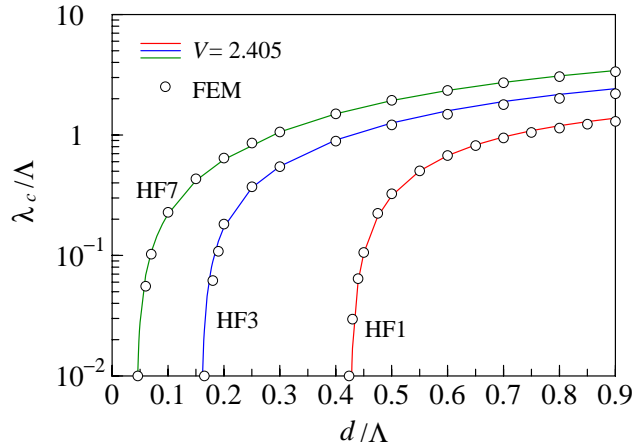

Fig. 1. Cut-off properties in a $\lambda_c/\Lambda$ versus $d/\Lambda$ phase-diagram for HF1, HF3, and HF7. The open data points are the FEM results while the solid curves show the predictions from Eq. (1) with $V(\lambda_c)=2.405$.

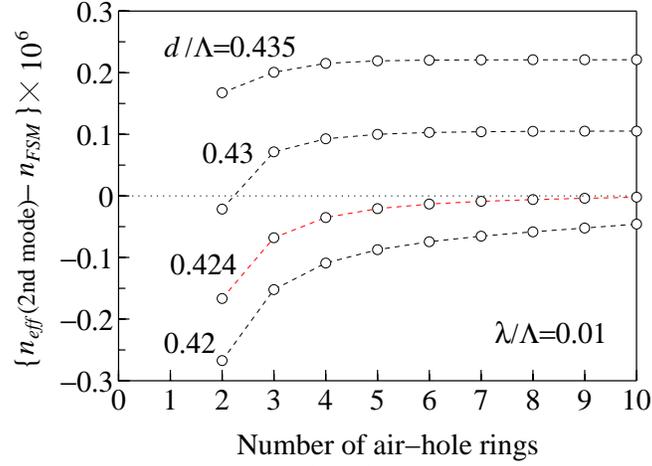
(a) HF1

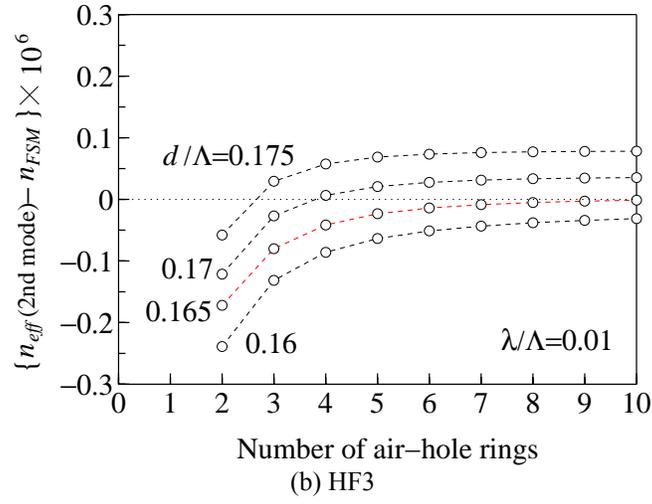
(b) HF3

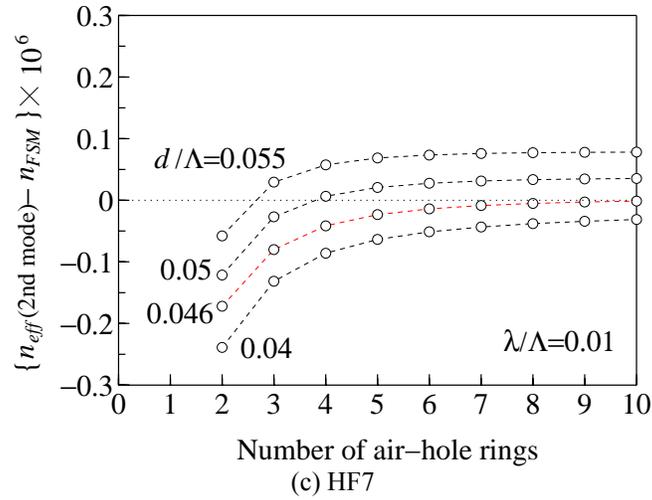
(c) HF7

Fig. 2. Impact of the number of air-hole rings on the value of $\Delta n = n_{eff}(\text{2nd mode}) - n_{FSM}$, where $n_{eff}(\text{2nd mode})$ is the effective index of the second order mode.

In order to justify the upper limit of the endlessly single-mode operation of the three types of HFs introduced in Table 1, in Fig. 2 we show the impact of the number of air-hole rings on the value of $\Delta n = n_{eff}$(2nd mode)$-n_{FSM}$, where $n_{eff}$(2nd mode) is the effective index of the second order mode. Keeping in mind that the definition of the cutoff wavelength is related to the value of $\Delta n = 0$, which is equivalent to the $W$ parameter being equal to zero (as the number of air-hole rings tend to infinity), we plot $\Delta n$ for various values of the normalized quantity $d/\Lambda$, for fixed operating wavelength of $\lambda/\Lambda = 0.01$. From the results in Fig. 2, we can clearly identify the appropriate limiting value of $d^*/\Lambda$, below which we have endlessly single-mode operation. Specifically, in Fig. 2(a), we see that, for HF1, this limiting value is $d^*/\Lambda$(HF1)=0.424, while from Fig. 2(b), $d^*/\Lambda$(HF3)=0.165, and from Fig. 2(c), $d^*/\Lambda$(HF7)=0.046.

From these results, we can define the $V$ parameter (the normalized frequency) and the $W$ parameter (the normalized transverse attenuation constant) for each holey fiber. Two slightly different definitions of $V$ parameters have been suggested recently, see Refs. [16], [17] and Refs. [18]-[20]. Here, we use the latter $V$ and $W$ parameter definitions

$$V = k a_{eff} \sqrt{n_{co}^2 - n_{FSM}^2}, \qquad (1)$$

$$W = k a_{eff} \sqrt{n_{eff}^2 - n_{FSM}^2}. \qquad (2)$$

We emphasized that this particular choice does not influence our overall conclusions. In these expressions, $k = 2\pi/\lambda$ is the free-space wave number, $a_{eff}$ is the effective core radius, $n_{co} = n_{silica} = 1.45$ is the core index, $n_{eff}(\lambda)$ is the effective index of the fundamental mode, and $n_{FSM}(\lambda)$ is the effective cladding index corresponding to the fundamental space-filling mode. In Ref. [18] it was found that by adjusting $a_{eff}$ to $\Lambda/\sqrt{3}$ for the HF1 the cut-off results in Fig. 1 are extremely well-described by $V(\lambda_c) = 2.405$ as for standard step-index fibers. For the other classes of HFs we likewise find that the values of $a_{eff}$ listed in Table 1 account well for the numerical FEM results in Fig. 1 as shown by the nice accounts for the FEM data points by the solid curves obtained from Eq. (1).

### 3. Susceptibility to longitudinal non-uniformities and bending loss properties

The susceptibility to longitudinal non-uniformities may be quantified by the coupling length $z_c$ between the fundamental mode and the fundamental space-filling mode [21] and using $n_{eff} + n_{FSM} \approx 2 n_{co}$ we have $z_c/\lambda \propto (\lambda/\Lambda)^{-2} W^{-2}$. Thus, when $W$ is increased the HF is less susceptible to longitudinal non-uniformities. Similar arguments apply for the leakage loss due to an air-hole cladding of finite extension [20]. The $V$ parameter reflects the same properties and in Fig. 3 we show the $V$ and $W$ values for endlessly single mode holey fibers (with $d/\Lambda = d^*/\Lambda$, see Table 1) as a function of $A_{eff}/\lambda^2$, where $A_{eff}$ is the effective mode area. From these curves we conclude that the HF1 is the one being less susceptible to longitudinal non-uniformities and the susceptibility in general increases for cores formed by an increasing number of removed air holes.

Next, in order to further support the above obtained general results, we consider the calculation of the macro-bending induced loss for the endlessly single-mode HF1 and HF3 with equivalent effective area of $A_{eff} = 200$ μm$^2$ at $\lambda = 1.55$ μm. The structural parameters are $\Lambda = 12.4$ μm and $d/\Lambda = 0.424$ for HF1 and $\Lambda = 7.0$ μm and $d/\Lambda = 0.165$ for HF3, respectively. We also show the macro-bending loss for LMA HF3 with $d/\Lambda = 0.25$ [8], but still with equivalent effective area of 200 μm$^2$ corresponding to a pitch $\Lambda = 7.85$ μm. We emphasize that the latter HF3 is not an endlessly single-mode fiber since $d/\Lambda = 0.25 > 0.165$. For calculation of the macro-bending induced loss we employ the tilted index model employed recently in Ref. [22]. The tilted index model is an exact mapping for a scalar description which is expected to work well for the LMA HFs with $\lambda \ll \Lambda$, see e.g. Refs. [23], [24]. Figure 4 shows the macro-bending loss property of HF1 and HF3 calculated through FEM. As clearly seen, HF1 has lower bending loss compared to the endlessly single-mode HF3 with $d/\Lambda = 0.165$, while HF3 with $d/\Lambda = 0.25$ has lower bending loss compared to HF1. The oscillatory characteristics of the HF1 bending loss originates from the coupling between the fundamental core mode and weakly localized cladding modes [25]. For HF3 the cladding has a significantly reduced air-hole diameter (compared to HF1) which does not support such weakly localized cladding modes. As a consequence we observe no oscillatory behavior for the endlessly single-mode HF3.

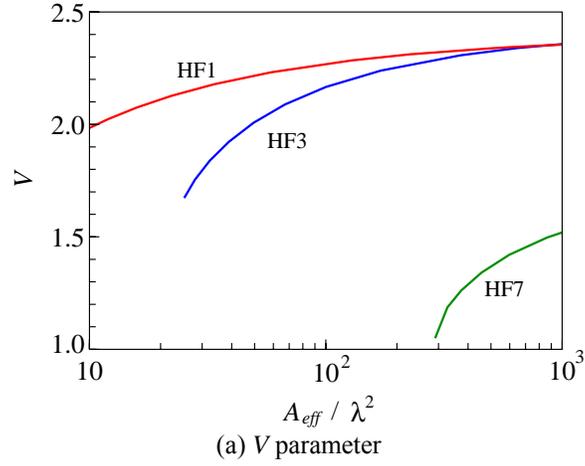

(a) *V* parameter

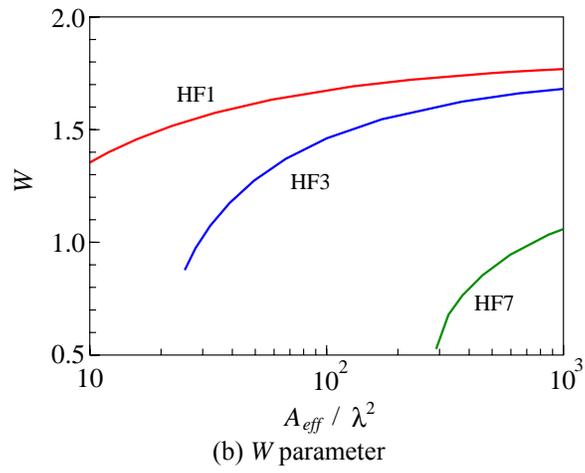

(b) *W* parameter

Fig. 3. Plot of the *V* and *W* parameters versus $A_{eff}/\lambda^2$ for air-hole diameters corresponding to the endlessly single-mode limit.

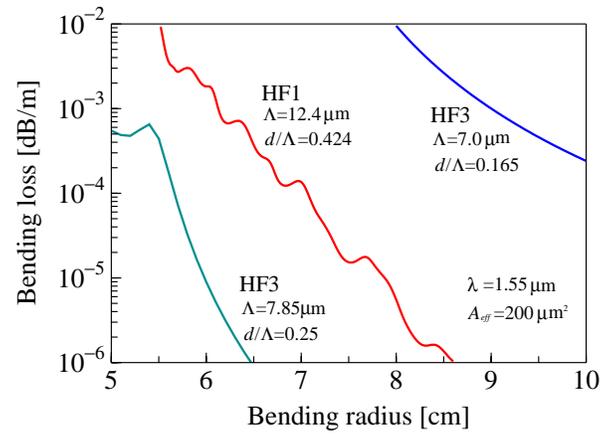

Fig. 4. Macro-bending loss for HF1 and HF3 with equivalent effective area of $A_{eff}$ = 200 μm² at λ = 1.55 μm. The structural parameters are Λ = 12.4 μm and d/Λ = 0.424 for HF1 and Λ = 7.0 μm and d/Λ = 0.165 for HF3, respectively. Macro-bending loss for HF3 with Λ = 7.25 μm and d/Λ = 0.25 is also plotted.

## 4. Conclusions

In conclusion we have investigated the cut-off properties of holey fibers for different core designs and we have accurately determined the single-mode and multi-mode phases as well as the air hole diameters limiting the endlessly single-mode regime. By comparing *V* and *W* parameters for the different core designs at the theoretical endlessly single-mode limit we conclude that in general HF1 is less susceptible to longitudinal non-uniformities compared to the other designs for equivalent effective areas. This conclusion relies heavily on the correct identification of the air-hole diameters limiting endlessly single-mode operation. In practice it is however well accepted that single-mode operation can be extended to somewhat higher air-hole diameters due to the high-loss nature of high-order modes. The practical limit depends quite a lot on the longitudinal quality of the fiber and the bending radius, but in e.g. Ref. [8] broad-band single-mode operation was experimentally observed with normalized air-hole diameters of 0.45 and 0.25 for HF1 and HF3, respectively.